\documentclass[12pt]{article}          

\usepackage{graphicx}
\usepackage{amsmath}
\usepackage{amssymb}
\usepackage{natbib}		

\begin{document}


\title{Coarse-grained simulation of polymer translocation through an 
artificial nanopore}

\author{Yves Lansac$^\ast$, Prabal K. Maiti$^\dagger$, and 
Matthew A. Glaser$^\ast$ \\
$^\ast$ Condensed Matter Laboratory, Department of Physics, and \\
Ferroelectric Liquid Crystal Materials Research Center, \\
University of Colorado, Boulder, CO 80309, USA \\
$^\dagger$ Materials and Process Simulation Center, \\
Division of Chemistry and Chemical Engineering, \\
California Institute of Technology,\\
Pasadena, CA 91125, USA}

\date{}                              

\maketitle

\begin{abstract}
The translocation of a macromolecule through a nanometer-sized pore is an 
interesting process with important applications 
in the development of biosensors for single--molecule analysis and in 
drug delivery and gene therapy.
We have carried out a molecular dynamics simulation 
study of electrophoretic translocation of a charged polymer 
through an artificial nanopore to explore the feasibility of
semiconductor--based nanopore devices for ultra--fast DNA sequencing. 
The polymer is represented by a simple bead--spring model designed 
to yield an appropriate coarse-grained description of the phosphate
backbone of DNA in salt--free aqueous solution. 
A detailed analysis of single translocation event is presented to assess 
whether the passage of individual ions through the pore can be detected by 
a nanoscale field--effect transistor by measuring variations in electrostatic 
potential during polymer translocation. We find that it is possible to 
identify single 
events corresponding to the passage of counterions 
through the pore, but that discrimination of individual ions on the 
polymer chain based on variations in electrostatic potential is problematic.
Several distinct stages in the translocation process are identified,
characterized by changes in polymer conformation and by variations in the
magnitude  and direction of the internal electric field induced by the 
fluctuating charge distribution. 
The dependence of the condensed fraction of counterions on Bjerrum length 
leads to significant changes in polymer conformation, which profoundly
affect the dynamics of electrophoresis and translocation. 
\end{abstract}

\clearpage


\section{Introduction}

In 1993, Branton and Deamer demonstrated that DNA could be threaded
through a biological pore and that, by measuring the variations in the
accompanying ionic current, information 
about DNA secondary structure could be obtained \citep{branton}. Although 
these results were very 
promising, little progress has been made towards actual sequencing of DNA 
due to the limited voltage range that can be applied across a biological pore 
and the difficulty in measuring the current variations, because the shot 
noise is comparable to the expected signal \citep{shahid}.
An alternative approach currently under investigation \citep{shahid}
consists in using
the new capabilities in nanoscale semiconductor
technology to create a more optimized pore geometry. 
A device consisting of a nanometer-sized pore in a thin silicon membrane  
with vertical transistors positioned along the wall of the pore may, in 
principle, be able to detect charge passing through the pore by measuring
image charges
in the transistor. By reducing the noise and maximizing the charge and 
current sensitivity, the proposed geometry may achieve DNA
sequencing with order of magnitude improvements in speed and cost and
minimal pre- and post-measurement procedures compared to current sequencing
techniques. The expected scientific and therapeutic benefits will be dramatic, 
for example enabling rapid routine screening for treatable genetic disorders.

Although an empirical approach based on trial and error can lead to some 
progress in the fabrication of such devices, a fundamental physical 
understanding of the delicate interplay between the physico-chemical behavior
of the macromolecule in solution and the interaction with the confining
geometry induced by the solid surface of the pore is highly desirable. 
Molecular modeling is a powerful tool
to investigate the collective behavior of systems resulting from the complex
interactions between their individual components.
Such an approach may allow us to gain insight into the phenomena of
driven polymer translocation through a nanometer-scale pore from both an applied
and a fundamental point of view. From an applied point of view, we can 
assess the feasibility of sensing and identifying individual
nucleotides, to aid in the design of semiconductor-based devices and
in the interpretation of the experimental measurements. In order to achieve
this goal, atomistic modeling using high accuracy molecular models must be used
to represent the properties of single stranded DNA, aqueous solution and 
substrate quantitatively. With this pre-requisite, it is possible
to study the local structure, conformation, and short time dynamics of 
single-stranded DNA in confined geometries, and to investigate the translocation
of ions and counterions across the channel as a function of the pore diameter 
and shape and of the magnitude of the potential difference applied. 
From a more fundamental point of view, it is of interest to study the mechanism 
of polymer translocation and the behavior of polyelectrolytes in confined
geometries. This is of relevance to the process of pore entry, including studies
of the role of counterion charge density gradients and of the
role played by hydrodynamic effects. 
A better understanding of this process can guide the design of pore 
geometries that effectively steer DNA molecules toward the pore.

Polymer translocation involves traversal of a free energy
barrier arising from several distinct physical effects, including loss of
conformational entropy of the macromolecule upon traversal (an
entropic barrier) \citep{muthu,muthu2}, mismatch in dielectric constant 
between the aqueous solution and the silicon pore (an electrostatic 
barrier) \citep{parsegian} and 
specific interactions of the macromolecule with the pore surface (an enthalpic 
barrier) \citep{zimm}. 
In this study, we use a coarse-grained approach, in which groups of atoms are 
represented by single interaction sites, to capture many of the 
essential physical features of the system. By dramatically decreasing 
the number of degrees of freedom present, this model can engender improved 
basic 
understanding of field-driven polymer translocation over long length and 
time scales.  

The remainder of this article is organized as follows:
In section II, we define the interaction potential used in this study,
the parametrization used to represent the phosphate backbone of DNA in 
salt--free aqueous solution, and the simulation methodology. 
In section III we present results obtained on the translocation
process and on the influence of Bjerrum length and external electric
field on the conformational characteristics of the polymer.
Finally, conclusions drawn from this study as well as future research directions
are discussed in section IV.

\section{Methodology}

\subsection{Interaction potential}

We study the translocation of a single negatively charged polymer chain  
designed to represent the phosphate backbone of a DNA fragment in 
an aqueous solution containing sodium counterions. Each molecule in this 
three-species mixture is represented as a set of spherical interaction sites. 
A coarse-grained approach in which groups of atoms are represented by single 
interaction sites is used in order to reduce the computational cost and to 
enable the study of the translocation process on long timescales.

We choose a simple interaction potential of the form:
$U({\bf r}^{N}) = U_{\rm str} + U_{\rm vdw} + U_{\rm coul} + U_{\rm ext} 
+ U_{\rm wall}$. 
The first term decribes intramolecular valence interactions, with $U_{\rm str}$ 
representing bond stretching interactions. We chose to model the phosphate
backbone as a flexible chain without bond angle bending
or dihedral torsion interactions in our model. It must be pointed out that the
chain is not fully flexible due to intramolecular van der Waals
and Coulombic interactions between second and more distant neighbors. 
The last four terms describe nonbonded interactions, 
with $U_{\rm vdw}$, $U_{\rm coul}$, $U_{\rm ext}$ and $U_{\rm wall}$
representing van der Waals, Coulombic, external electric field, and 
molecular site -- substrate interactions respectively. 
The pore is defined as a cylindrical channel
(with axis along the $z$ direction) in a solid substrate, which is treated
as a smooth van der Waals surface, and the external electric field $E$ driving
polymer translocation is in the $+z$ direction. 
These contributions to the total potential are defined as:

\begin{eqnarray}
U_{\rm str} & = & \sum_{{\rm bonds} \atop ij} 
\frac{1}{2} k_{r} {(r_{ij} - r_{\rm eq})}^{2} \\ \nonumber \\
U_{\rm vdw} & = & {\sum_{i<j}}^{'} 4 {\epsilon}_{ij}
\left [  { {\left ( {\frac{{\sigma}_{ij}}{r_{ij}} }\right )}^{12}
- {\left ( {\frac{{\sigma}_{ij}}{r_{ij}} }\right )}^{6} } \right ]
\label{vdw} \\ \nonumber \\
U_{\rm coul} & = & {\sum_{i<j}}^{'} \frac{q_{i} q_{j}}{r_{ij}} \label{coul}
\\ \nonumber \\
U_{\rm ext} & = & {\sum_{i}} E q_{i} z_{i} \\ \nonumber \\
U_{\rm wall} & = & {\sum_{i}} 4 {\epsilon}_{iw}
\left [  { {\left ( {\frac{{\sigma}_{iw}}{r_{iw}} }\right )}^{12}
- {\left ( {\frac{{\sigma}_{iw}}{r_{iw}} }\right )}^{6} } \right ]
\label{wall} 
\end{eqnarray}

The primes on the sums in Eqs. (\ref{vdw}) and (\ref{coul}) indicate that
nearest--neighbor intramolecular nonbonded interactions 
are excluded from the
sums. The distance $r_{ij}$ between sites $i$ and $j$ is defined as 
$r_{ij} = |{\bf r}_{ij}| = |{\bf r}_j - {\bf r}_i|$ where ${\bf r}_i$ is the
position of site $i$ and $z_i$ its vertical component.


The remaining step in building the molecular model is fixing the interaction
parameters. These parameters need to be provided for every distinct
combination of site types. In the next subsection, the site types and the
interaction parameters used in the case of the coarse-grained model will be 
specified.

\subsection{Parametrization of the coarse-grained model}

An appropriate parametrization of the coarse-grained molecular model is
obtained via an approximate mapping from the corresponding physical system.
The DNA backbone is negatively charged and the concentration of 
positively--charged sodium ions is such that the system is neutral overall.

Water molecules are treated as point van der Waals (Lennard-Jones) 
interaction sites, with the Lennard-Jones (LJ) parameters determined from
a `corresponding states' mapping of the triple point of water onto that
of the LJ system \citep{johnson} yielding ${\sigma}_{w} = 0.29$ nm and 
${\epsilon}_w = 0.79$ kcal/mol.
These values are taken as the unit of length ${\sigma}_0$ and the unit of 
energy ${\epsilon}_0$.  The mass of a single water molecule is the unit 
of mass, $m_0$ = 18 a.u. Using these values, a unit of time $t_0 = 
\sqrt{m_{0} {{\sigma}_0}^2 / {\epsilon}_{0}} \simeq 7 \times 10^{-10} s$, 
a unit of charge $q_0 = \sqrt{{\epsilon}_{0} {\sigma}_{0}} \simeq  e / 12$,
and a unit of electric field $E_0 = {\epsilon}_{0} / (q_{0} {\sigma}_{0}) 
\simeq 1400$ V/$\mu$m are defined, where $e$ is the magnitude of the 
electron charge.

The sodium ions are treated as point interaction sites with the same LJ
parameters as water and with a reduced charge $q_{Na^+} = \alpha q_0$. We
have investigated systems with $\alpha = 1.2$, $2.4$ and $6$ corresponding
to a charge $q_{Na^+} = e /10$, $e / 5$ and $e / 2$. 
These values have been chosen in order to take into account dielectric
screening effects in an approximate way.  For simplicity, the mass of sodium 
ions is set equal to the mass of water molecules.
 
The phosphate backbone consists of alternating phosphate and deoxyribose
groups. Each group is represented by a van der Waals point interaction 
site (Figure~\ref{fig:dna}) with van der Waals parameters  
${\sigma}_{\rm ph} = 2 {\sigma}_0$ and ${\epsilon}_{\rm ph} = {\epsilon}_0$. 
The phosphate sites carry reduced electric charge 
$q_{\rm ph} = - q_{\rm {Na^+}}$ while the deoxyribose sites do not carry 
any electric charge. The bond stretching parameters are set 
to the following values: $r_{\rm eq} = {\sigma}_0$ and $k_{r} = 24$ kcal/mol.
The phosphate chain is quite flexible although
not fully flexible due to the van der Waals and Coulombic 
interactions between second and more distant neighbors. 
The masses of both phosphate and deoxyribose sites are set to 5 $m_0$. 
Geometrical combination rules for $\sigma$ and $\epsilon$ were applied
to parametrize the LJ interactions between unlike species.

The pore is defined as a cylindrical channel in a solid substrate, which 
is treated as a smooth van der Waals surface, the atomic sites interacting
with the nearest point on the surface of the pore or substrate. For 
simplicity, the LJ parameters of the wall sites are the same as those of 
water. The diameter of the pore is $d_{\rm pore} = 5 {\sigma}_0$
with a length $l_{\rm pore} = 2 d_{\rm pore}$.
The diameter of the pore is $2.5$ times larger than that of the polymer
in order to accomodate both the polymer chain and at most one solvation 
sphere. In Branton's experiments, the alpha--hemolysin pore diameter
has been estimated to be between $18$ to $26 \AA$ \citep{branton,meller} 
which is $2$ to $2.5$ times the diameter of a single-stranded DNA 
fragment ($D_{ds} = 20 \AA$, $D_{ss} = 10 \AA$ \citep{tinland}). The length 
of the pore
was estimated to be $l_{\rm pore} = 52 \AA$, in good agreement with the value
chosen in our simulation. Reduced variables defined in this
subsection are used in the remainder of this article.

\subsection{Simulation methodology}

We carry out NVT (constant particle number, volume, and temperature) 
molecular dynamics simulations of a single charged polymer 
in solution, under an external field, with periodic boundary conditions.
The unit cell contains a reservoir separated by an impenetrable
substrate of thickness $l_{\rm pore}$ and surface area equal to the lateral 
dimension 
of the unit cell, and containing a cylindrical hole (nanopore)
of diameter $d_{\rm pore}$ (Figure~\ref{fig:cartoon}).
    
The chain consists of 40 sites, alternating between phosphate and 
deoxyribose groups, and the initial state is built such that the charged 
polymer is in the reservoir outside the pore cavity. The system contains, in
addition to the polymer, 6000 water molecules and 20 Na$^+$ counterions 
with initial positions chosen randomly in the unit cell.  
The simulation was carried out at a reduced density ${\rho}^* = 0.85$ 
and a reduced temperature $T^* = k_{B}T / {\epsilon}_0 = 0.75$,
corresponding to the dense fluid phase of a LJ system.
The applied external electric field of magnitude  $E^\ast = E / E_{0} = 0.5$,
corresponding to a field $E = $709 V/$\mu$m, 
is turned on after an initial equilibration run of duration $t^\ast = 500$.
In the experiments conducted
with the alpha - hemolysin nanopore, a voltage of 70-300 mV was applied 
across the system. The electrodes are located centimeters away from the
pore, but the bulk of the voltage drop occurs across the biological membrane, 
leading to an estimated electric field of 15-60 V/$\mu$m, at least an order of
magnitude smaller than the field used in the present work.
Such a large field is used in the present study in order to ensure that
field-driven translocation occurs on a time scale accessible to simulations.

A single-timestep velocity-Verlet MD integration scheme was used to integrate
the equations of motion with a timestep $\delta t = 2.5 \times 10^{-3}$, and 
the weak-coupling algorithm of Berendsen {\sl et al.} \citep{weak_coupling}
was used to maintain constant temperature. All LJ interactions were truncated
at 2.5 ${\sigma}_0$ and no long-range corrections were applied.
Long-range Coulomb interactions are evaluated to high accuracy
using the particle-mesh Ewald (PME) method \citep{darden,essmann}.
The PME technique, which is derived from the conventional Ewald method 
\citep{ewald,toukmaji}, makes use of the FFT to efficiently compute
the long range part of the electrostatic interaction.
In this study we used a highly optimized version of the PME with a relative
accuracy of $10^{-4}$. 

We have carried out 3 simulations of systems with reduced charge magnitude 
$q^\ast = 1.2$, 2.4 and 6.0 for durations of $4.2 \times 10^6$, $3.0 \times
10^6$ and  $5.2 \times 10^6$ timesteps, respectively.

\section{Results}

The translocation process consists of three distinct stages: In the first 
stage,
the polymer drifts through the reservoir, driven by the external electric 
field. In the second stage, the polymer is pushed against the lower substrate
surface and behaves like a 2--dimensional chain wandering in search of the
pore entrance. In the third stage, the polymer finds the entrance to the 
pore and translocation takes place.

Due to the relatively small system size, the time
spent by the polymer in the second stage is small compared to that in 
the two other stages. We first study
the electrophoresis of the charged polymer in the reservoir, then the 
kinetics of
the translocation process and the possibility of identifying 
individual ions passing through the pore from the electrostatic signal  
produced.

\subsection{Electrophoresis in the reservoir}

Electrophoresis is a complex process dependent on the electrophoretic
friction coefficient \citep{manning81}. This coefficient arises both from
the macroion/solvent
interactions, comprising the monomer/solvent interaction (the intrinsic
friction) and from hydrodynamic interactions between monomer pairs, 
and from macroion/ion interactions. Since the macroion is negatively
charged, it is surrounded by an oppositely charged liquid atmosphere 
\citep{fuoss}. The
force acting on the ions in the atmosphere is transmitted to the solvent,
and therefore the atmospheric liquid behaves as a charged volume.
Under an applied electric field, this charged liquid is subject to a volume
force in the opposite direction from the drift velocity of the macroion.
The macroion thus moves against a local hydrodynamic flow that slows 
its motion (relative to a hypothetical drift without the charged
liquid). In addition, the electric field deforms 
the charged volume, increasing the atmospheric charge density at the end
of the polymer opposite to the end which drift and decreasing ahead of
the macroion. This effect induces an internal field (often referred in the
literature as the relaxation field, or more accurately, the asymmetry field) 
opposing the external electric field, which also slows down 
the macroion motion. 

Counterion condensation is in large part responsible for the structure and 
behavior of the liquid atmosphere and, as recognized in the late 60's 
independently by Oosawa and Manning \citep{oosawa,manning}, plays a crucial 
role in the strong attractive
interaction that acts between highly charged macroions such as DNA, leading
to significant conformational changes. Counterion condensation results in
a competition between Coulombic energy and entropy in minimizing the free
energy of an aqueous solution containing mobile ions in the vicinity of
an isolated macroion. For rodlike objects, whether the entropy 
or the Coulombic attraction dominates depends on the charge density 
of the macroion measured in terms of the Bjerrum length 
${\lambda}_{B} = q^2/ k_{B}T$, defined 
as the distance at which the Coulombic interaction
between two charges is equal to the thermal energy $k_{B}T$.
Counterion condensation occurs when the distance $b$ between charges
on the macroion is small enough for the dimensionless parameter
$\Gamma = {\lambda}_{B} / b$ (the Manning parameter) to exceed unity. 
With the reduced units introduced in the previous section, a reduced
Bjerrum length ${\lambda}_{B}^\ast = {q^\ast}^2 / T^\ast$ is defined.
In the present model, the distance between 2 charges is $b = 2 {\sigma}_0$,
leading to a Manning parameter $\Gamma = {q^\ast}^2 / 2 T^\ast$. 
Values for ${\lambda}_{B}^\ast$ and $\Gamma$ are listed in Table~\ref{table}
for the systems with reduced charge $q^\ast = 1.2$, $2.4$, $6.0$.
At the reduced temperature chosen in this study, Coulombic interactions
dominate over the thermal energy for $q^\ast = 6.0$
and $q^\ast = 2.4$ systems, which thus should exhibit strong counterion 
condensation.
For the $q^\ast = 1.2$ system, $\Gamma = 0.96$, which is close enough to 
unity to also exhibit counterion condensation.
However, in this case we expect to have a much less strongly bound
counterion - macroion complex than for $q^\ast = 6.0$ and $q^\ast = 2.4$.

It is difficult to define an unambiguous criterion for condensation. Indeed,
as pointed by Oosawa \citep{oosawa}, for a coiled polymer chain each charged
group makes a sharp and deep potential hole at its position, each linear
part of the chain makes a sharp and deep potential valley along its length
and the coiled chain as a whole makes a potential trough in its apparent
volume. Counterions located in these 3 regions are considered
to be condensed, with the counterions at charged group holes being localized
and the counterions in the other two regions being mobile.
Therefore, an estimate of the degree of condensation  
depends on the criterion used to discriminate between 
free and condensed counterions. In the present study, a counterion is said 
to be condensed if it is within a distance $r^{\ast}_c$ from an ion of
the polymer. 

Figure~\ref{fig:snapshots} shows representative configurations for the three  
systems both with and without applied external electric field $E^\ast$. 
We first study the system behavior without an external applied field. 
The three systems exhibit counterion condensation but the nature of the
resulting macroion--counterion complex varies 
with the magnitude of the charge. For $q^\ast = 6.0$, 
a large majority of the counterions are closely associated with the
polymer chain, being mostly bound or in the narrow potential valley
along the length of the chain, while for $q^\ast = 1.2$ the counterions 
are mostly 
located in the potential valley of the polymer chain. An intermediate
situation occurs for $q^\ast = 2.4$.
Figure~\ref{fig:gmc_comparison} shows the radial 
pair correlation function $g_{ic}(r^{\ast})$ between ions on the chain
and counterions, for $q^{\ast} = 1.2$, 2.4 and 6.0 and $E^\ast = 0$. 
The common features exhibited are a main peak at $r^{\ast} \simeq 1.5$, 
corresponding to localized condensed counterions, and
a shoulder around $r^{\ast} \simeq 2.0$, corresponding to 
mobile counterions in the narrow potential valley around the chain. 
The height of the main peak for $q^\ast = 6.0$ is roughly 7 times larger 
than for $q^\ast = 1.2$ and 4 times larger than for $q^\ast = 2.4$,
confirming a stronger counterion localization
for larger charge magnitude. 
For $q^\ast = 1.2$,  Figure~\ref{fig:gmc} shows the relative importance
of longer range correlations with respect to short range correlations
with the presence of secondary peaks around $r \simeq 2.3 {\sigma}_0$ and 
$r \simeq 3.5 {\sigma}_0$, indicating 
that a significant counterion fraction is located in the potential trough 
of the chain.
Table~\ref{table} displays the average
number of condensed counterions for the 3 systems for   
different values of $r^{\ast}_c$ corresponding 
roughly to the bottom of the main peak in $g_{ic}$ ($r^{\ast}_c = 1.8$),
the shoulder ($r^{\ast}_c = 2.2$) and a value taking into
account the long range correlations ($r^{\ast}_c = 6.0$).

Under the applied external field, the counterions remain  
located very close to the macroion forming bound dipoles for $q^{\ast} = 6.0$.
In addition, a conformational change occurs with the formation of a 
collapsed polymer structure.
By contrast, for $q^\ast = 1.2$, the counterions are not condensed anymore
(Figure~\ref{fig:snapshots}).
The force produced by the applied external field overcomes the
attractive Coulombic force and strips the counterions from the
chain. As a first approximation, the reduced critical charge at which the
Coulombic force exactly balances the force produced by the external field is
$q^{\ast}_{c} = E^\ast {\bar r}^{\ast \; 2}$ where ${\bar r}^{\ast}$ is the 
average reduced distance between a polymer ion and a counterion. Assuming 
a condensed state with all counterions bound to 
the polymer, the estimate ${\bar r}^{\ast} \simeq
1.5$ -- $2.0$ leads to $q^{\ast}_{c} = 1.125$ -- $2.0$. 
Thus, no counterions will be stripped from the chain for $q^\ast = 6.0$,
while a large counterion fraction will be stripped away for
$q^\ast = 1.2$. In the latter case, $q^{\ast}_c$ is a lower limit 
since the counterions are not closely bound to the chain
(see Figure~\ref{fig:snapshots}). 
A Monte Carlo simulation study of the counterion condensation on a 
spherical macroion has shown the same phenomenon under a strong external
field \citep{tanaka}. With this estimate, the counterion fraction located in
the potential trough of the chain will be stripped from the polymer for 
$q^\ast = 2.4$. 

Figure~\ref{fig:gmc} shows the ion--counterion pair correlation function 
when an external electric field is applied on the systems.
For $q^\ast = 1.2$, the magnitude of the main peak for $q^\ast = 1.2$
is reduced by an order of magnitude, confirming that a large 
counterion fraction is stripped from the chain.
The electric field has a much less drastic effect on the two other systems: 
A small fraction of counterions are stripped from the chain
for $q^{\ast} = 2.4$ while for $q^\ast = 6.0$, the field enhances slightly
the counterion binding to the polymer. Average values for the number of 
condensed counterions under external electric field are given in
Table~\ref{table}. 

The main qualitative features exhibited during
the time evolution are independent of the chosen value for
$r^{\ast}_c$. Figure~\ref{fig:condense} shows the time evolution of the 
number of condensed counterions for the three systems for $r^{\ast}_c = 1.8$ 
and $r^{\ast}_c = 6.0$.
The $r^{\ast}_c = 6.0$ system does not
exhibit any significant changes in the number of condensed counterions
over the duration of the simulation. The counterions are mostly
condensed on the chain, and the electric field slightly enhances
this effect.
By contrast, for the $q^\ast = 1.2$ system, a strong decrease in the 
number of condensed counterions occurs in an external field, the new 
equilibrium configuration being reached after a time $t^{\ast} \simeq 2500$.
The $q^\ast = 2.4$ system exhibits an initial decrease
of the number of counterions induced by the external electric field similar
to that for $q^\ast = 1.2$ with a new equilibrium 
state also reached after a duration $t^{\ast} \simeq 2500$. This effect 
is weaker, however (see Table~\ref{table} and Figure~\ref{fig:gmc}) due to 
the fact that a larger fraction of 
counterions are condensed on the chain ($q^{\ast}_c < q^\ast = 2.4$). 
The behavior during the translocation process is discussed in the next 
subsection.

We turn now to the study of the polymer conformations as a function of 
magnitude of the site charge $q^\ast$ and electric field $E^\ast$. 
A global measure of the polymer conformation is 
the mean square radius of gyration $R_{g}^2$, defined as: 

\begin{equation}
R_{g}^2 = \frac{1}{N} \sum_{i = 1}^N <{({\bf r}_i - {\bf R})}^2>
\end{equation}

where ${\bf r}_i$ is the position of the $i^{\rm th}$ polymer site, $N$ is 
the number of sites in the chain, and $\bf R$ is the position of the center 
of mass of the chain.
Average values of $R_g^2$ over a duration $t^\ast = 2500$ for the 3 systems
with and without applied electric field are listed in Table~\ref{table}.
Figure~\ref{fig:gyration} shows the time evolution of $R_g^2$ 
for $q^\ast = 1.2$, 2.4 and 6.0.
For $q^{\ast} = 1.2$, the chain becomes more elongated under the action of the
external field because the mobile counterions are stripped away by
the field,  
leaving the bare negative ions on the polymer and leading to an electrostatic
stiffening of the polymer.  
For $q^\ast = 2.4$, the first effect of the electric field, before 
translocation through the pore, is to strip away 
the fraction of counterions which are in the potential trough of the polymer,
resulting in a loss of electrostatic screening of the negative charges on
the polymer and an increase in
$R_g^2$. Then the polymer chain is driven through the reservoir by the 
electric field and pushed against the lower substrate 
(Figure~\ref{fig:snapshots}). 
In this regime, the polymer behaves like a 2-dimensional chain and its radius 
of gyration may be significantly reduced \citep{foo97,foo98}. 
Conformational changes during the translocation process are discussed in the 
next subsection.

For $q^{\ast} = 6.0$, the radius of gyration is of the same order of magnitude 
as for $q^\ast = 1.2$ and $q^\ast = 2.4$ when no external field is applied.
Under an external electric field, no significant change in the magnitude of the
radius of gyration is observed over a time interval  $t^\ast \simeq 5000$, then
a strong decrease in $R^2_g$ occurs, indicative of the tightly collapsed 
structure adopted by the chain (Figure~\ref{fig:gyration}). 
Additional studies are needed to determine whether the observed collapse
is  induced by the external field or is simply caused by strong counterion 
condensation. Biological processes and theoretical works tend to favor the
latter hypothesis. 
The chain collapse is analogous to the packing of DNA into a cell 
\citep{bloomfield}. This
packing requires overcoming an enormous Coulombic barrier in a highly dilute 
aqueous solution containing a small concentration of polvalent cations.
Simulations and theory have demonstrated short range
attraction between two macroions modeled as charged cylinders  
\citep{gronbech,lyubartsev} or stiff polymers \citep{stevens99}. 
This attraction has been ascribed to correlated fluctuations 
of the counterions induced by counterion condensation \citep{ha,ray}. 
Self-attraction have been shown to occur also in flexible polyelectrolytes
\citep{kremer,pincus,brilliantov}.
Theoretical work using simple scaling arguments have shown that counterion
condensation modifies the second virial coefficient of a polyelectrolyte due 
to the fact that at low enough temperature (or at high enough counterion 
valence)
the counterions approach close enough to the macroion to form dipoles,
leading to charge - dipole and dipole - dipole interactions \citep{pincus}. 
Similar ideas have been developed by Brilliantov {\sl et al.} 
\citep{brilliantov}, who also predict 
also a first order phase transition between a stretched polyelectrolyte 
and a strongly collapsed polyelectrolyte as a function
of magnitude of electric charges due to counterion condensation. 

\subsection{Translocation process}

The thermodynamics of electrophoresis in the presence of a narrow pore is  
complex and involves traversal of entropic and enthalpic barriers
~\citep{muthu,muthu2,apell,slonkina,lubensky,sung,boehm,sebastian,kumar,lee}. 
When the chain enters 
the pore, its conformational entropy is reduced, due to chain elongation.
On the other hand, due to the small size of the
pore aperture a significant fraction of the counterions are stripped from the
chain and gain configurational entropy. The counterion unbinding also
results in an increase in the Coulombic
energy. In addition, the electrostatic energy of interaction of the macroion 
with the external field decreases as the negatively charged 
chain moves through the pore, and the electrostatic interaction energy of 
the free counterions also decreases. 
The situation is reversed when the polymer exits the pore. However, 
the gain in conformational entropy of the polymer upon exiting the pore
is probably greater than
the entropy loss associated with pore entry, since before 
pore entry the polymer is pushed against the
lower substrate surface ($z = - z_0$) and behaves like a conformationally
restricted two-dimensional
chain while it behaves like a three-dimensional chain upon exiting the pore.  
Detailed free energy calculations are needed to study the
importance of these contributions in the translocation process.
We focus here on the kinetics of the translocation process under a large
electric field.

Translocation was only observed in the $q^\ast = 2.4$ simulation.
As we have seen in the previous subsection, for $q^\ast = 6.0$, strong 
counterion condensation results in repulsive screened Coulomb 
interactions between the ions on the chain and self--attraction due to 
charge--dipole, dipole--dipole interactions and/or charge fluctuation along 
the chain.
In the $q^\ast = 6.0$ system, the chain adopts a collapsed structure with an 
effective diameter larger
than the pore diameter, effectively preventing translocation.
There is, {\sl a priori}, no reason for the $q^\ast = 1.2$ system not to 
translocate since the chain adopts an extended conformation.
The simulation has to be performed over a 
long enough period of time to permit the polymer to find the pore entrance
via diffusion along the substrate surface. 

Due to the large external field, translocation of the polymer in the 
$q^\ast = 2.4$ system occurs after a
short drift time spent in the reservoir. The polymer finds the entrance of
the pore after a time interval $\Delta t^\ast = 4000$ ($\Delta t = 2.8$ ns) 
after application of the field. 
The upper part of Figure~\ref{fig:translocation} shows several 
stages in the translocation process. For clarity, only the polymer
chain and the counterions are shown.
The translocation starts when the first bead (head) enters the pore at
$z = -z_0$ and ends when the last bead (tail) exits the pore at $z = z_0$.
The total translocation process occurs over a period of time 
$\Delta t^\ast \simeq 950$, corresponding to 0.665 ns. 
Timescales reported for translocation of ions and water molecules 
are in the range $10^{-6} - 10^{-9}s$, in reasonable agreement with our 
results. 
However this time is much shorter than the characteristic time reported for 
ssDNA translocation ($10^{-6}$ s) \citep{branton,meller,henrickson} due to 
the larger electric field used in the present work. 

We have decomposed the translocation process into three stages \citep{slonkina}
in which the
conformation of the polymer and its interactions with the pore are
significantly different: the first stage 
corresponds to the translocation of the head from the entrance to the exit of
the pore, the second stage corresponds to the exit of the head from the pore
and the translocation of the tail from the reservoir to the entrance of the 
pore, and the third stage corresponds to the translocation of the tail from 
the entrance to the exit of the pore (see Figure~\ref{fig:translocation}). 
The duration of each stage of translocation is:
$\Delta t^{\ast}_1  =  t^{\rm head}_{z = z_0} - 
t^{\rm head}_{z = -z_0} = 145.74$ ($\Delta t_1 = 0.102$ ns);
$\Delta t^{\ast}_2  =  t^{\rm tail}_{z = -z_0} - 
t^{\rm head}_{z = z_0} = 424.75$ ($\Delta t_2 = 0.297$ ns) and
$\Delta t^{\ast}_3  =  t^{\rm tail}_{z = z_0} - 
t^{\rm tail}_{z = -z_0} = 379$ ($\Delta t_3 = 0.266$ ns).
The duration of the second and third stages are comparable, but are nearly 
three times as long as the first stage.
These differences in time are probably due to variation in the magnitude of 
the total electric field $E^{\ast}_{\rm total} = E^\ast + E^{\ast}_{\rm int}$ 
experienced by the polymer during the translocation, where
$E^{\ast}_{\rm int}$ is the internal field induced by the charge
distribution. 

The bottom part of Figure~\ref{fig:translocation} shows the 
reduced electrostatic potential $V^\ast = V / V_0$ ($V_0 = {\epsilon}_0
/ q_0$) due to ions and counterions in the 
$(xz)$ plane passing through the middle of 
the pore. We clearly see an inversion in the sign of the potential difference
across the pore during the second stage of polymer translocation
due to a change in the
relative charge density on the two substrate surfaces.
This change leads to an inversion in the direction of the internal field 
$E^{\ast}_{\rm int}$, which in turn slows down the translocation 
process.
The upper and lower substrate surfaces 
behave like a capacitor, with an excess of positive charges on the
upper surface -- the free counterions trying to translocate and pushed against
the surface by the external electric field -- and an excess of negative charges 
on the lower surface -- the polymer chain wandering at the pore entrance. 
This situation is reversed when enough negatively charged monomers have 
translocated, analogous to the discharge of a capacitor. 

Figure~\ref{fig:internal_field} shows quantitatively the time evolution 
of the reduced internal electric field $E^{\ast}_{\rm int}$, defined as 
$E^{\ast}_{\rm int} = \Delta V^{\ast}_{\rm pore} / l^{\ast}_{\rm pore}$, 
across the pore during the translocation event, with 
$\Delta V^{\ast}_{\rm pore}$ being the 
average potential difference across the pore defined as
$\Delta V^{\ast}_{\rm pore} = 1/4 \sum_{i = 1}^{4} (V^\ast(x_i,y_i,-z_0)
- V^\ast(x_i,y_i,z_0))$ with 
($x_i = 0$; $y_i = \pm r_0$) for $i = 1,2$ and ($x_i = \pm r_0$; $y_i = 0$)
for $i = 3,4$ (see Figure~\ref{fig:cartoon}).
It confirms that the internal field decreases in magnitude and changes 
sign when the
middle of the chain crosses the middle of the pore (i.e., in the second 
stage) resulting in a smaller magnitude of total electric field 
$E^{\ast}_{\rm tot}$ and a slower translocation process. 

We turn now to the identification of single translocation events involving
either a counterion or charged sites on the polymer.
In order to mimic the signal recorded at the gate of a transistor located on 
the inner cylindrical surface of the pore, we have computed the electrostatic 
potential $V^\ast$ at 4 equidistant positions ($x = 0$; $y = \pm r_0$) and
($x = \pm r_0$; $y = 0$) on the surface of the mid--plane ($z = 0$)
of the pore. 
We have simultaneously recorded events corresponding to the passage of
charged sites (either counterions or charged polymer beads) through the $z = 0$
plane to investigate the correlations between signal and site positions. 
Figures~\ref{fig:potential} and \ref{fig:potential2} show the time evolution 
of $V^\ast$ averaged over the four sensors as well as ion passage 
events during the translocation process. 
A positive spike records the passage of a charge in the
same direction as the external electric field while a negative spike
records the passage of a charge in the opposite direction. No counterions are 
pulled into the pore with the polymer so we do not have any counterions
translocating against the electric field direction during polymer translocation.
A strong correlation is found between counterion positions and the signal
(sharp peaks in $V^\ast$). It is also fairly easy to detect when the polymer 
translocation takes place (and its approximate duration), but the 
discrimination of individual charges on the polymer chain is more problematic
due mainly to the small separation between neighboring charged sites on the
chain, fluctuations in chain position, and the simultaneous
translocation of counterions. Further analysis or filtering of the signal
is required to achieve single-ion discrimination even with this simple model. 

We notice that a significant larger number of counterions find the entrance 
of the pore during the third stage of polymer translocation.
During the translocation process, the number of condensed counterions on the
chain increases significantly (Figure~\ref{fig:condense}). 
Due to the applied electric field, an excess counterion 
density is located near the upper substrate surface and a significant fraction 
of them -- the ones near the aperture -- condense back onto the polymer 
as soon as the head of the chain exits the nanopore 
(see Figure~\ref{fig:translocation}). 
The condensed counterions near the tail of the polymer are relatively mobile
and have a significant probability of being
stripped from the chain due to the external field and friction with 
the water molecules.
The translocation of these counterions is facilitated by the polymer, 
which acts as a guide to the pore entrance. 

In addition, the head of the polymer exiting the nanopore during the 
translocation process experiences a stronger electric field than inside 
the pore (see above).
The polymer tail is still pushed by the field against the lower 
substrate surface creating an anchoring point on the substrate 
(see Figure~\ref{fig:translocation}) 
while the head of the chain is stretched by the electric field
at the pore exit. This effect results in an increase in $R_g^2$ 
(Figure~\ref{fig:gyration}). After the head has exited the pore, counterion 
condensation back on the head, screening the Coulombic
interactions and reducing the stretching force, as well as a weaker 
anchoring at the entrance of the pore due to the presence of fewer monomers, 
are responsible of the decrease in polymer extension, the chain behaving like 
a rubber band. As the translocation proceeds, 
the tail enters the pore and the release of the anchoring point induces a
significant contraction of the chain.
After the translocation, the chain behaves as a three-dimensional chain, 
resulting in an increase in $R_g^2$.

\section{Conclusions}

Using a coarse-grained bead--spring model, we have studied the electrophoretic
translocation of a charged polymer through a nanopore
for various charge magnitudes under the application of an external electric 
field. Three regimes can be identified, corresponding to drift of
the polymer in the reservoir, diffusion of the polymer along the substrate 
surface in
search of the pore entrance and polymer translocation through the pore.
The three systems studied, with reduced charges, $q^\ast = 1.2$, 2.4 and 6.0,
respectively, exhibit counterion
condensation when no electric field is applied, but the $q^\ast = 1.2$
system presents a much less strongly bound macroion--counterion complex than 
the others two. Counterion condensation leads to significant changes in
polymer confomation, with a rather extended chain conformation for 
$q^\ast = 1.2$
and a collapsed chain conformation for $q^\ast = 6.0$, due to self-attraction
induced by charge--dipole/dipole--dipole interactions and/or counterion
fluctuations along the chain. An external electric field can strip the
counterions from the chain, depending on the relative magnitude of the field
and the charge of the ions/counterions, inducing significant changes in  
polymer conformation.  
Due to the strong external electric field used, translocation occurs
quickly for the $q^\ast = 2.4$ system and simulation over longer
time scales should lead to the translocation of the $q^\ast = 1.2$ system.
By contrast, translocation in the $q^\ast = 6$ system is inhibited due to 
the collapsed conformation adopted by the chain. 
The kinetics of the polymer translocation are not constant over the whole
process and the chain translocates more slowly when its midpoint
crosses the mid-plane of the pore due to an inversion in the direction of the
internal electrostatic field produced by the time-varying distribution of
charges. The upper and lower substrate surfaces behave qualitatively
like a capacitor which undergoes a discharge. 
By recording the electrostatic potential we were able to identify the passage 
of individual counterions unambiguously. However, it is difficult to identify 
translocation events involving single charged sites on the polymer due mainly
to the small distance between neighboring ions, fluctuations
in polymer position, and simultaneous counterion translocation.

This work is mostly exploratory and many questions remain unanswered.
Future work will focus on both the translocation phenomenon and on  
bulk electrophoresis.
Using the simple model presented in this work, we will study the 
electrophoresis process in detail, including the effect of external
electric field on counterion condensation as a function of the ion charge
magnitude and macroion concentration. Special emphasis will be 
given to the modification of macroion conformation induced by the external 
field.
In the framework of the simple model introduced
in this study, we will study the translocation process statistically by
steering the polymer through the pore and performing
free energy computations to assess the relative importance of 
entropic and enthalpic contributions to the free energy of pore traversal.  
Since one of our aims is to study the feasibility of a biosensor for fast
DNA sequencing we have initiated large-scale atomistic simulations of ssDNA
in aqueous solution to study the electrostatic field
produced by different nucleotides in the vicinity of the 
macromolecule, to explore the possibility of discriminating 
different bases based on their electrostatic signature.

%
%

%
%

\clearpage

\begin{figure}[htb]
\caption{ 
Mapping of the phosphate backbone (left) onto a coarse-grained model.
The phosphate group is represented by a spherical interaction site (red)
carrying a charge $q = - q_{\rm {Na^+}}$ and the deoxyribose group is
represented by a spherical interaction site (green) carrying no charge.
}
\label{fig:dna}
\end{figure}

\begin{figure}[htb]
\caption{
Side view (left) and top view (right) of the simulated system, comprised of a 
reservoir with a substrate containing a nanopore of radius $r_0$ and length 
$2 z_0$. 
}
\label{fig:cartoon}
\end{figure}

\begin{figure}[htb]
\caption{
Representative configurations of the 3 polyelectrolyte systems, without 
external field (top) at $t^\ast = 500$ and with applied electric field 
(bottom) at
$t^\ast = 10500$ for $q^\ast = 1.2$ and $q^\ast = 6.0$ and at $t^\ast = 4000$ 
(before translocation) for $q^\ast = 2.4$. For clarity, the water molecules
have not been displayed.
}
\label{fig:snapshots}
\end{figure}

\begin{figure}[htb]
\caption{ 
Ion-counterion  pair distribution function $g_{ic}(r)$ for
$q^\ast = 1.2$ (solid line), 2.4 (dashed line) and 6.0 (dot--dashed line)
without external field.
}
\label{fig:gmc_comparison}
\end{figure}

\begin{figure}[htb]
\caption{
Ion-counterion pair distribution function $g_{ic}(r)$ for
$q^\ast = 1.2$, 2.4 and 6.0 (from top to bottom) for $E^\ast = 0$ (solid line)
and $E^\ast = 0.5$ (dashed line).
}
\label{fig:gmc}
\end{figure}

\begin{figure}[htb]
\caption{ 
Time evolution of the number of condensed counterions for $q^\ast = 1.2$,
$2.4$ and $6.0$ (from top to bottom) using a condensation criterion
$r^{\ast}_c = 1.8$ (thick solid line) and $r^{\ast}_c = 6.0$ (thin solid 
line). For $q^\ast = 2.4$, the arrows 
represent the time location of the four translocation stages (A, B, C, D)
presented on Figure~\ref{fig:translocation}.
}
\label{fig:condense}
\end{figure}

\begin{figure}[htb]
\caption{
Time evolution of the mean square radius of gyration $<R^2_g>$ for
$q^\ast = 1.2$, 2.4 and 6.0 (top to bottom).
For $q^\ast = 2.4$, the arrows
represent the time location of the four translocation stages (A, B, C, D)
presented on Figure~\ref{fig:translocation}.
}
\label{fig:gyration}
\end{figure}

\begin{figure}[htb]
\caption{
Top: Four stages in the translocation process, from left to right: 
head of the polymer at the entrance of the pore ($t^\ast = 4472.75$), 
head at the exit of the pore ($t^\ast = 4618.5$), tail at the entrance 
($t^\ast = 5043.25$), tail at the exit ($t^\ast = 5422.25$).
Bottom: Corresponding reduced electrostatic potential $V^\ast = V / V_0$
($V_0 = {\epsilon}_0 / q_0$) 
produced by the charge distribution in the plane ($xz$) passing through
the center of the pore. Values of the potential outside the range chosen
(-5,+5) are assigned to the colors corresponding to the minimum and maximum
values, and for clarity the water molecules have not been represented.  
}
\label{fig:translocation}
\end{figure}

\begin{figure}[htb]
\caption{ 
Time evolution of the induced electric field $E^\ast$ produced by the 
distribution
of charges. The arrows represent the time location 
of the four translocation stages presented on Figure~\ref{fig:translocation}. 
}
\label{fig:internal_field}
\end{figure}

\begin{figure}[htb]
\caption{ 
Time evolution of the reduced electrostatic potential $V^\ast$
and of the events corresponding to the passage of counterions (top) and ions 
 (bottom) through the midplane of the pore ($z = 0$), during polymer 
translocation.
A positive peak corresponds to an ion or counterion passing in the same 
direction as the field while a negative peak corresponds to an ion or 
counterion passing in opposite direction. 
}
\label{fig:potential}
\end{figure}

\begin{figure}[htb]
\caption{ 
Time evolution of the reduced electrostatic potential $V^\ast$ and of 
the events corresponding to the passage of ions 
at $z = 0$, during the polymer translocation. 
Magnification of Figure~\ref{fig:potential}.
}
\label{fig:potential2}
\end{figure}

\clearpage

\begin{table}[htb]
\begin{center}
\caption{Electrostatic and conformational properties of the simulated systems.}
\label{table}
\begin{tabular}{|c|c|c|c|} \hline \hline
 & $q^\ast = 1.2$ & $q^\ast = 2.4$ & $q^\ast = 6.0$\\ 
\hline
${\lambda}_{B}^\ast$ & 1.92 & 7.68 & 48.0\\ \hline
$\Gamma$ & 0.96 & 3.8 & 24.0 \\ \hline
Number of & 2.347 $\pm$ 0.158 ($r_c^{\ast} = 1.8$) &
4.761 $\pm$ 0.712 (1.8) & 11.785 $\pm$ 1.897 (1.8) \\ \cline{2-4}
condensed counterions & 2.897 $\pm$ 0.263 ($r_c^{\ast} = 2.0$) &
5.706 $\pm$ 1.074 (2.0) & 12.726 $\pm$ 2.044 (2.0) \\ \cline{2-4}
($E^{\ast} = 0$)& 10.174 $\pm$ 2.266 ($r_c^{\ast} = 6.0$) &
12.358 $\pm$ 2.428 (6.0) & 16.866 $\pm$ 1.648 (6.0) \\ \hline
Number of & 0.283 $\pm$ 0.138 ($r_c^{\ast} = 1.8$) &
3.029 $\pm$ 0.902 (1.8) & 14.254 $\pm$ 0.165 (1.8) \\ \cline{2-4}
condensed counterions & 0.268 $\pm$ 0.149 ($r_c^{\ast} = 2.0$) &
3.623 $\pm$ 1.265 (2.0) & 14.510 $\pm$ 0.079 (2.0) \\ \cline{2-4}
($E^{\ast} = 0.5$)& 1.567 $\pm$ 0.494 ($r_c^{\ast} = 6.0$) &
10.404 $\pm$ 3.158 (6.0) & 18.026 $\pm$ 0.049 (6.0) \\ \hline
$< R^2_g >$ ($E^{\ast} = 0$) & 23.030 $\pm$ 1.375 & 25.789 $\pm$ 1.459
& 20.202 $\pm$ 1.008 \\ \hline
$< R^2_g >$ ($E^{\ast} = 0.5$) & 60.007 $\pm$ 2.716 & 60.837 $\pm$ 7.157 
& 13.961 $\pm$ 0.279 \\ \hline \hline
\end{tabular}
\end{center}
\end{table}

%
%

\clearpage
\setcounter{figure}{0}

\begin{figure}[htbp]
\includegraphics[width=4.5in]{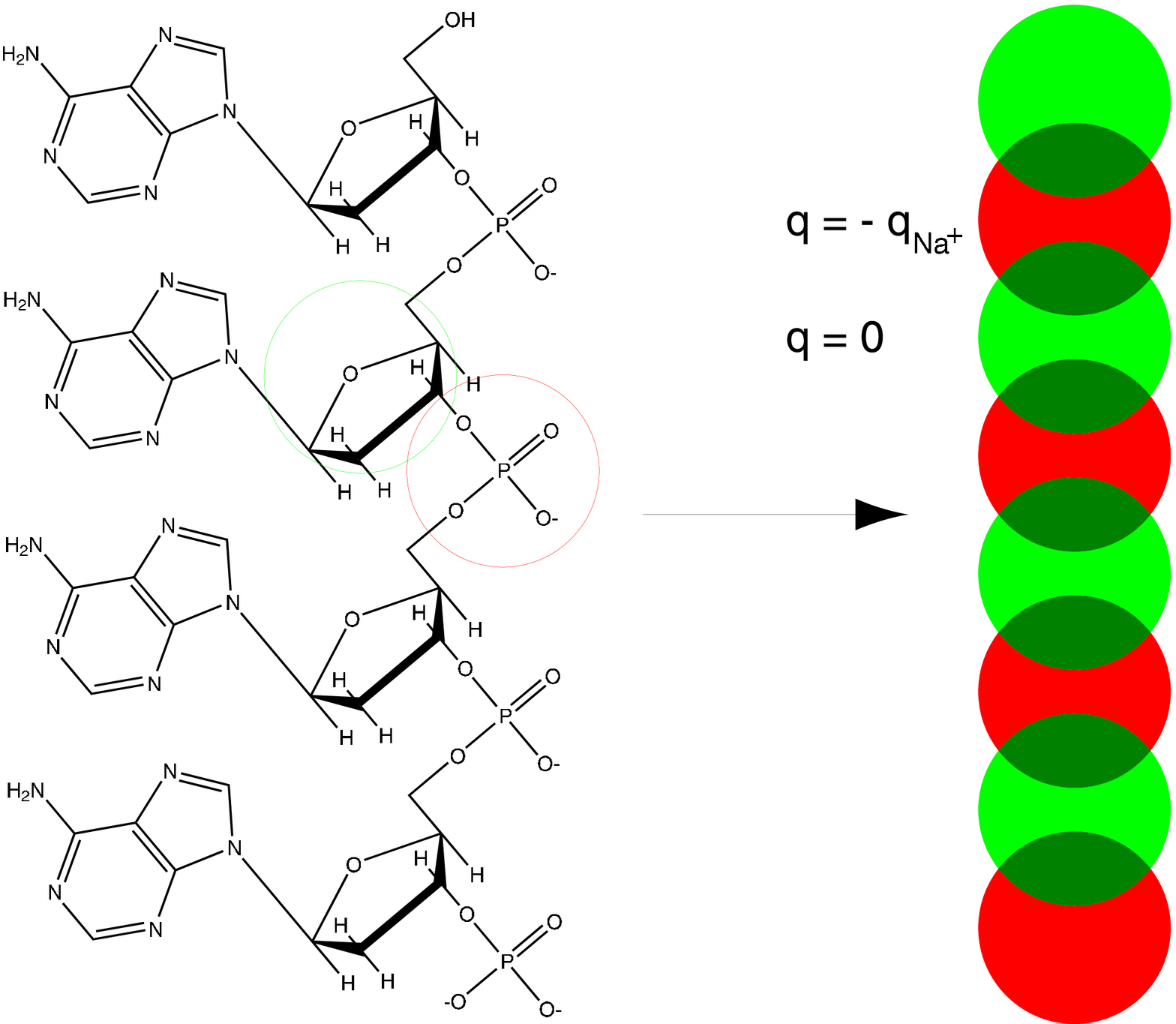}
\caption{ }
\end{figure}

\newpage

\begin{figure}[htbp]
\includegraphics[width=6in]{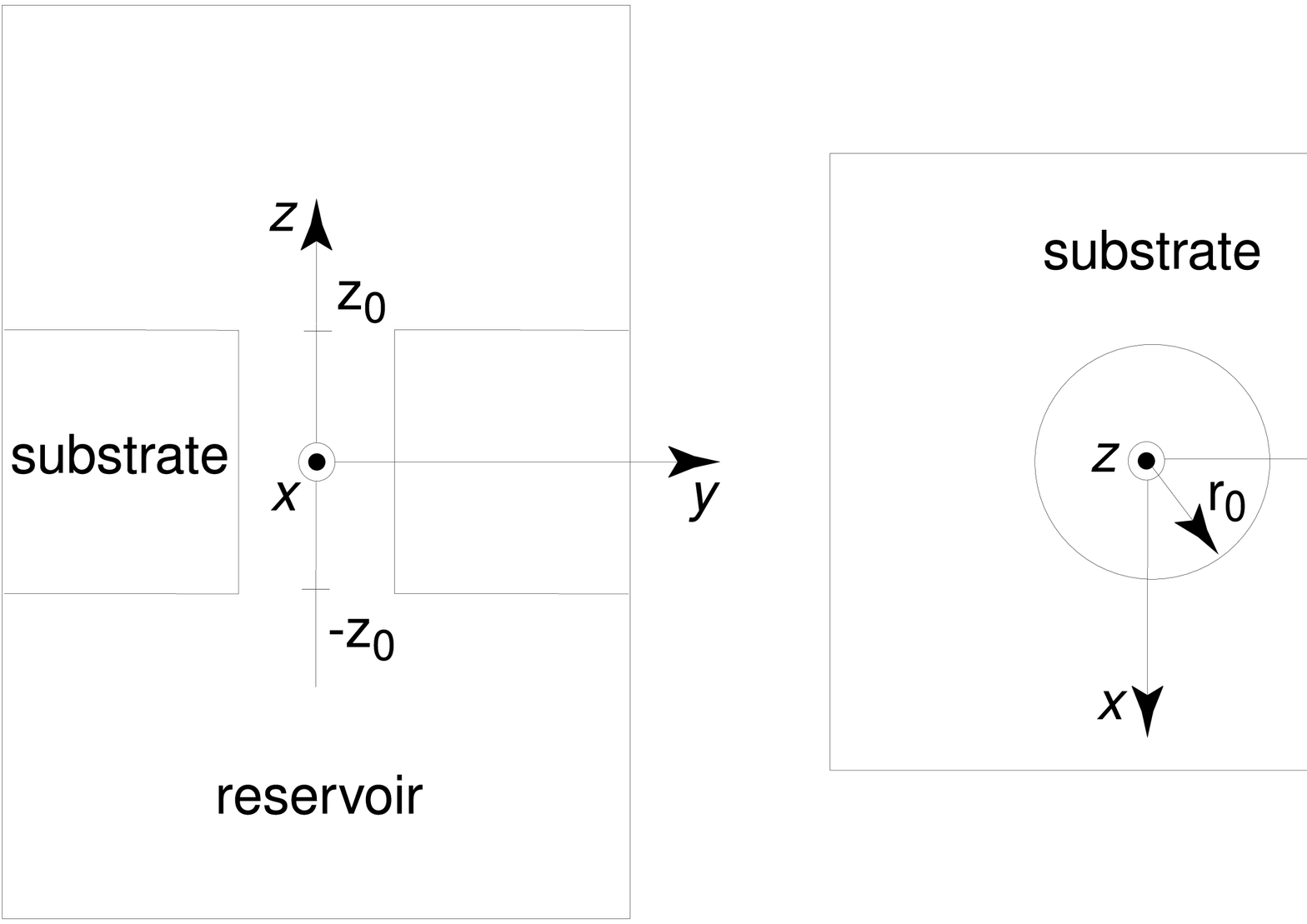}
\caption{ }
\end{figure}

\newpage

\begin{figure}[htbp]
\includegraphics[width=6.0in]{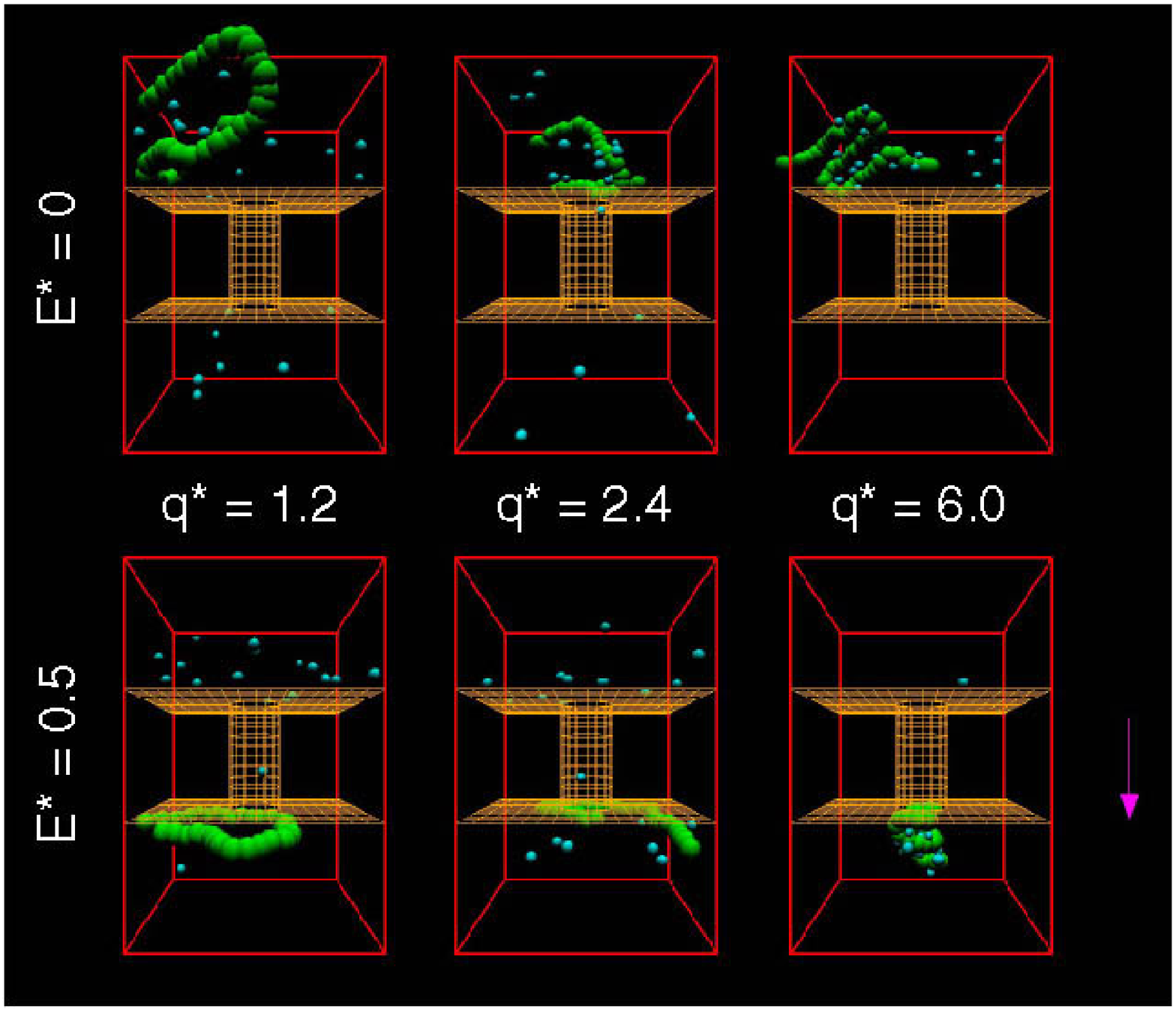}
\caption{ }
\end{figure}

\newpage

\begin{figure}[htbp]
\includegraphics[width=6in]{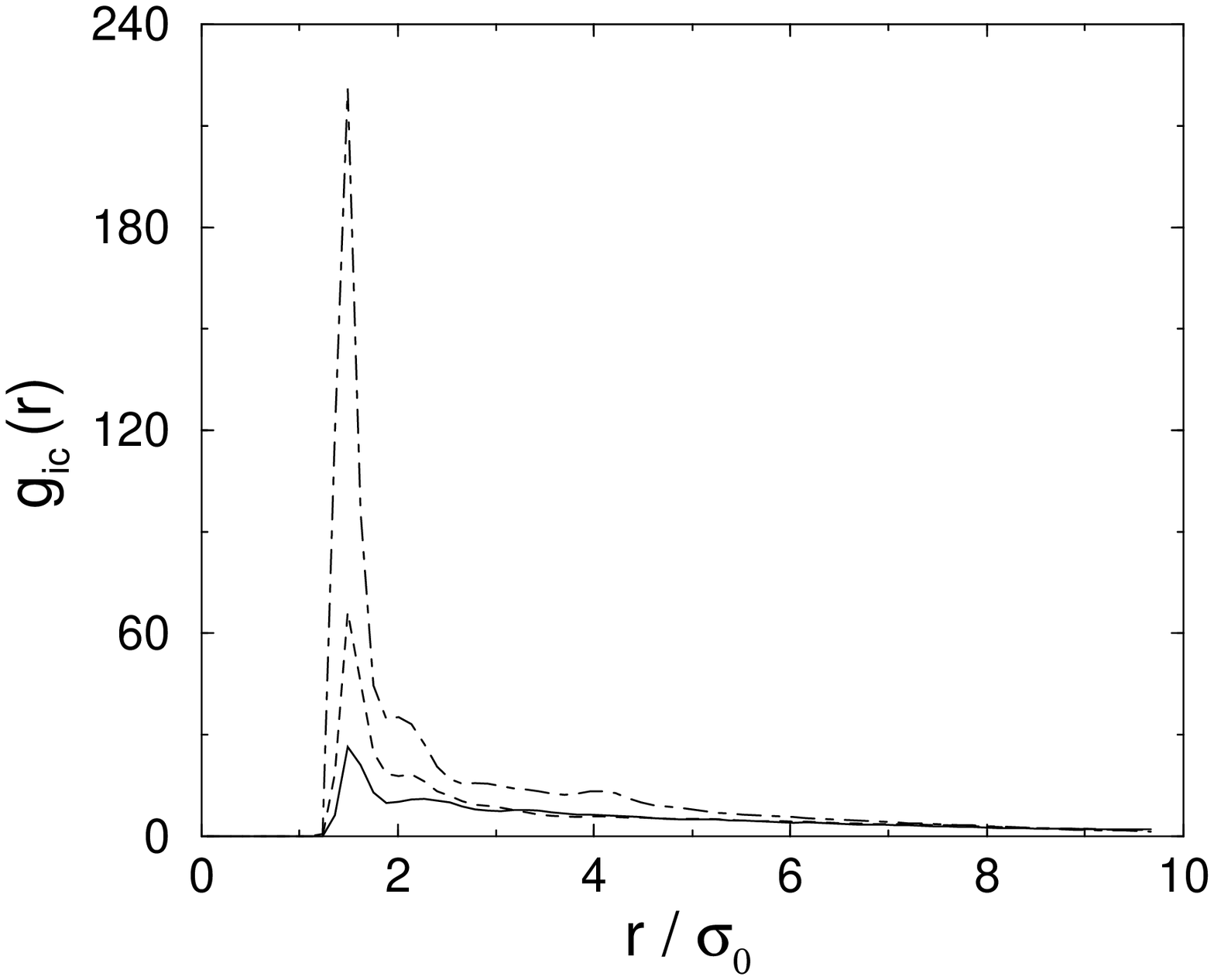}
\caption{ }
\end{figure}

\newpage

\begin{figure}[htbp]
\includegraphics[width=5.5in]{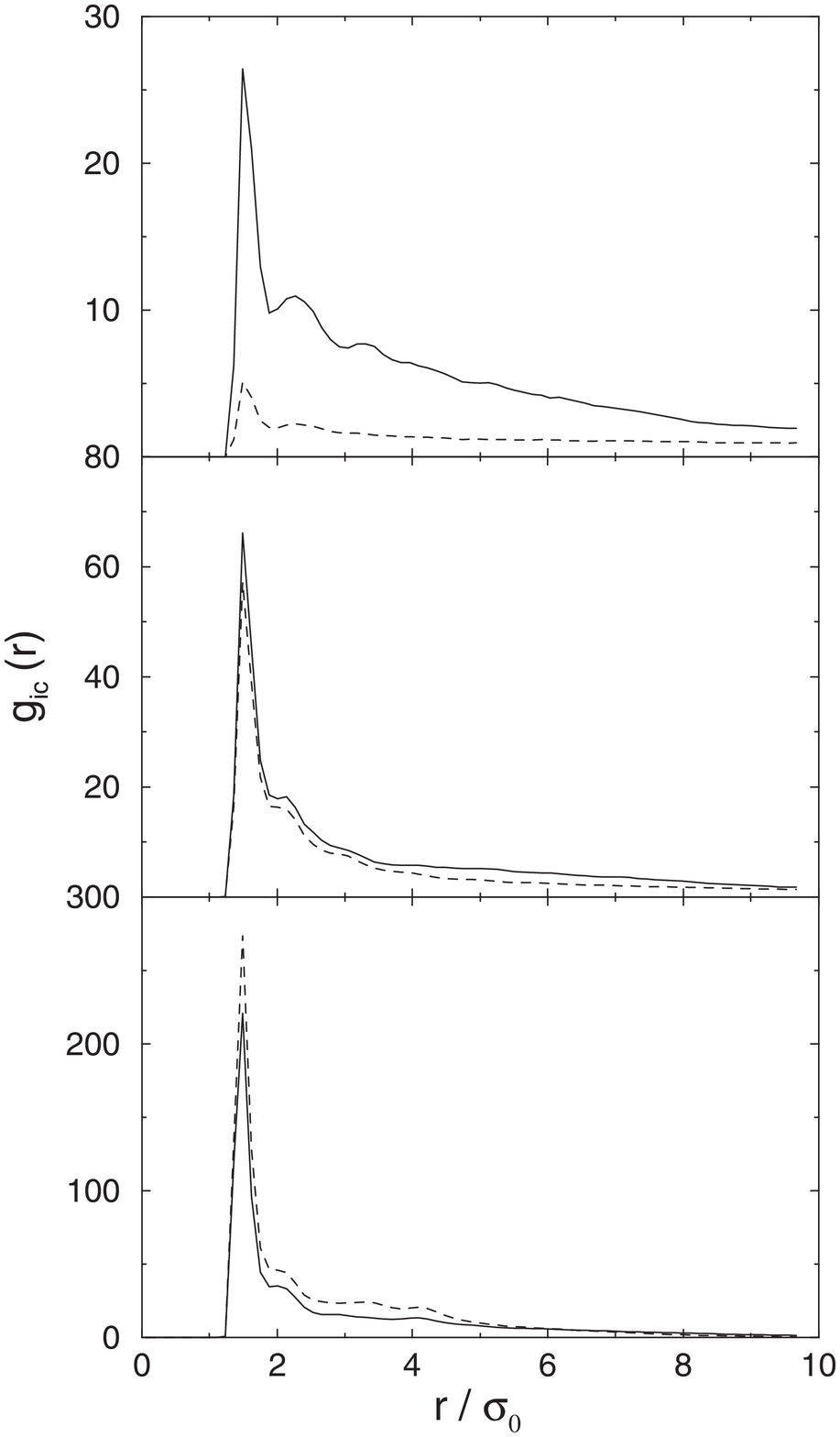}
\caption{ }
\end{figure}

\newpage

\begin{figure}[htbp]
\includegraphics[width=4.0in]{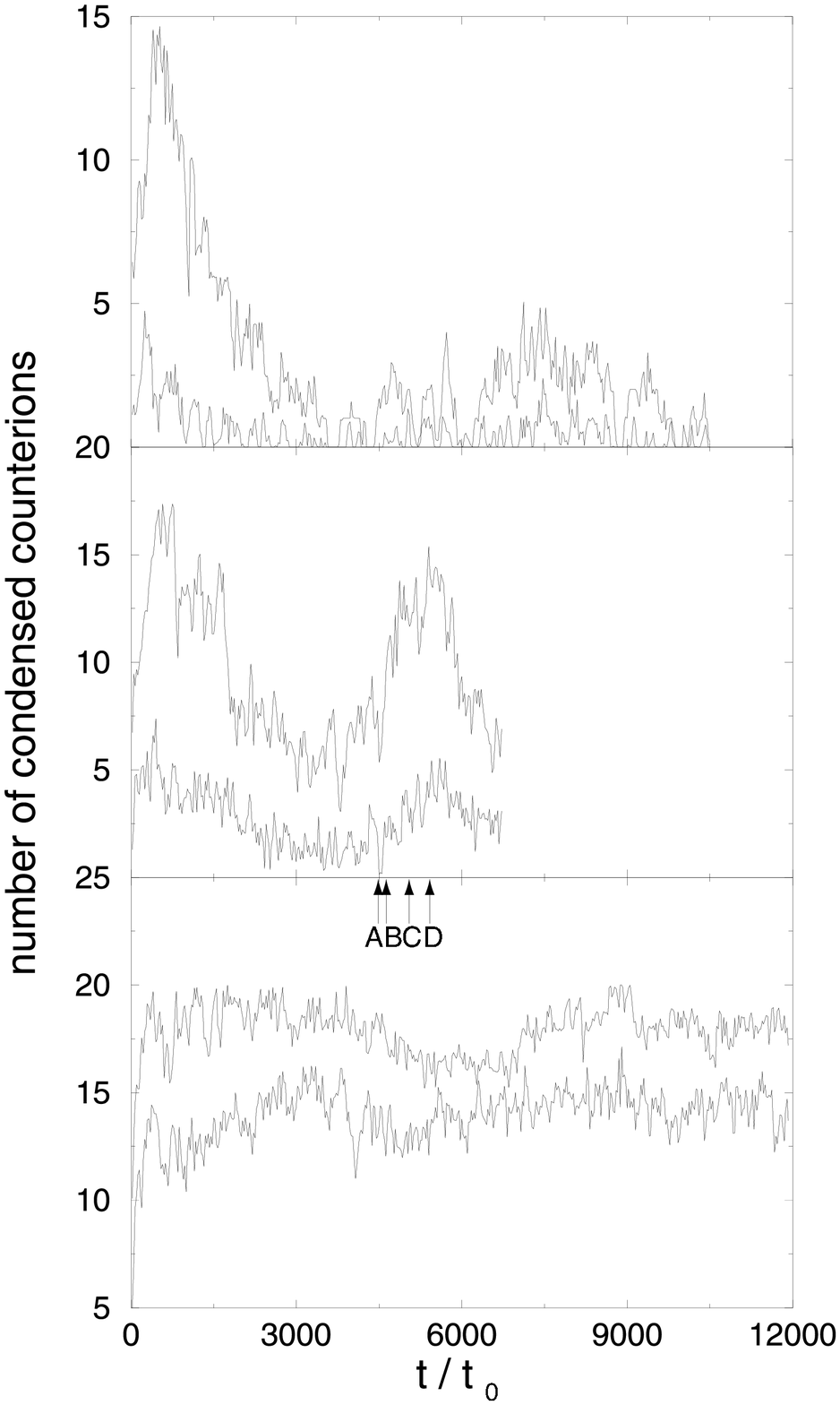}
\caption{ }
\end{figure}

\newpage

\begin{figure}[htbp]
\includegraphics[width=5.0in]{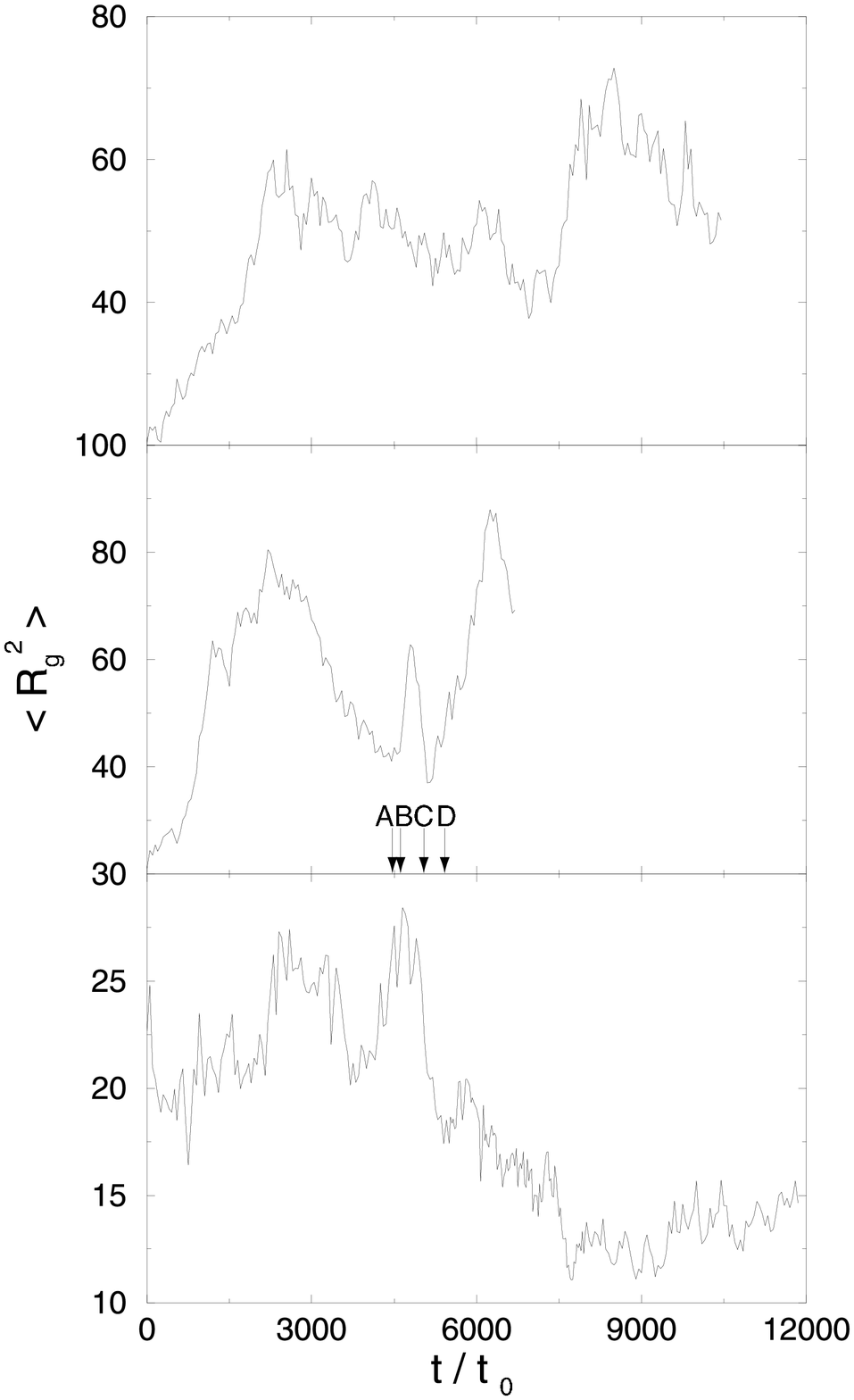}
\caption{ }
\end{figure}

\newpage

\begin{figure}[htbp]
\includegraphics[width=6.5in]{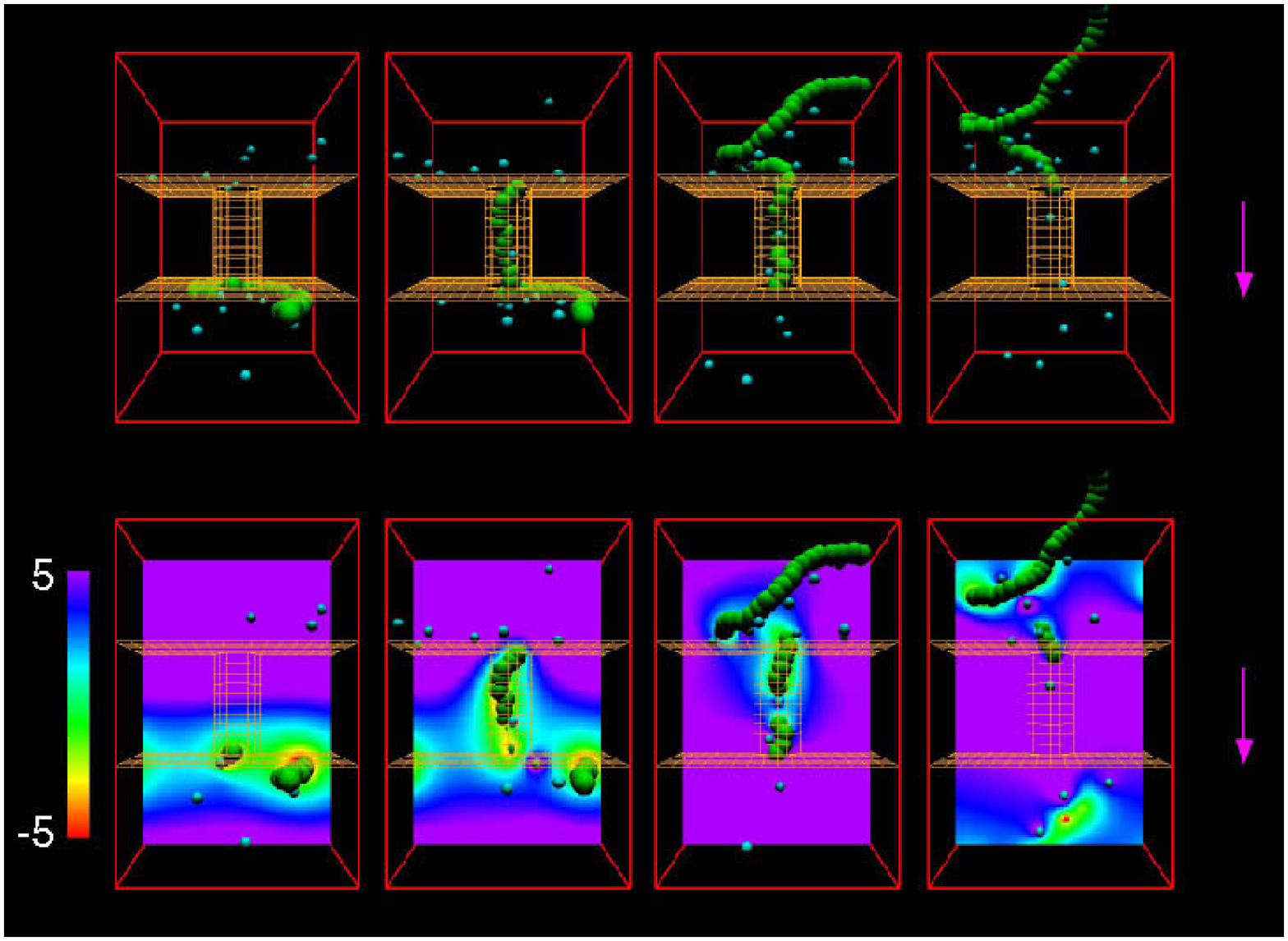}
\caption{ }
\end{figure}

\newpage

\begin{figure}[htbp]
\includegraphics[width=5.5in]{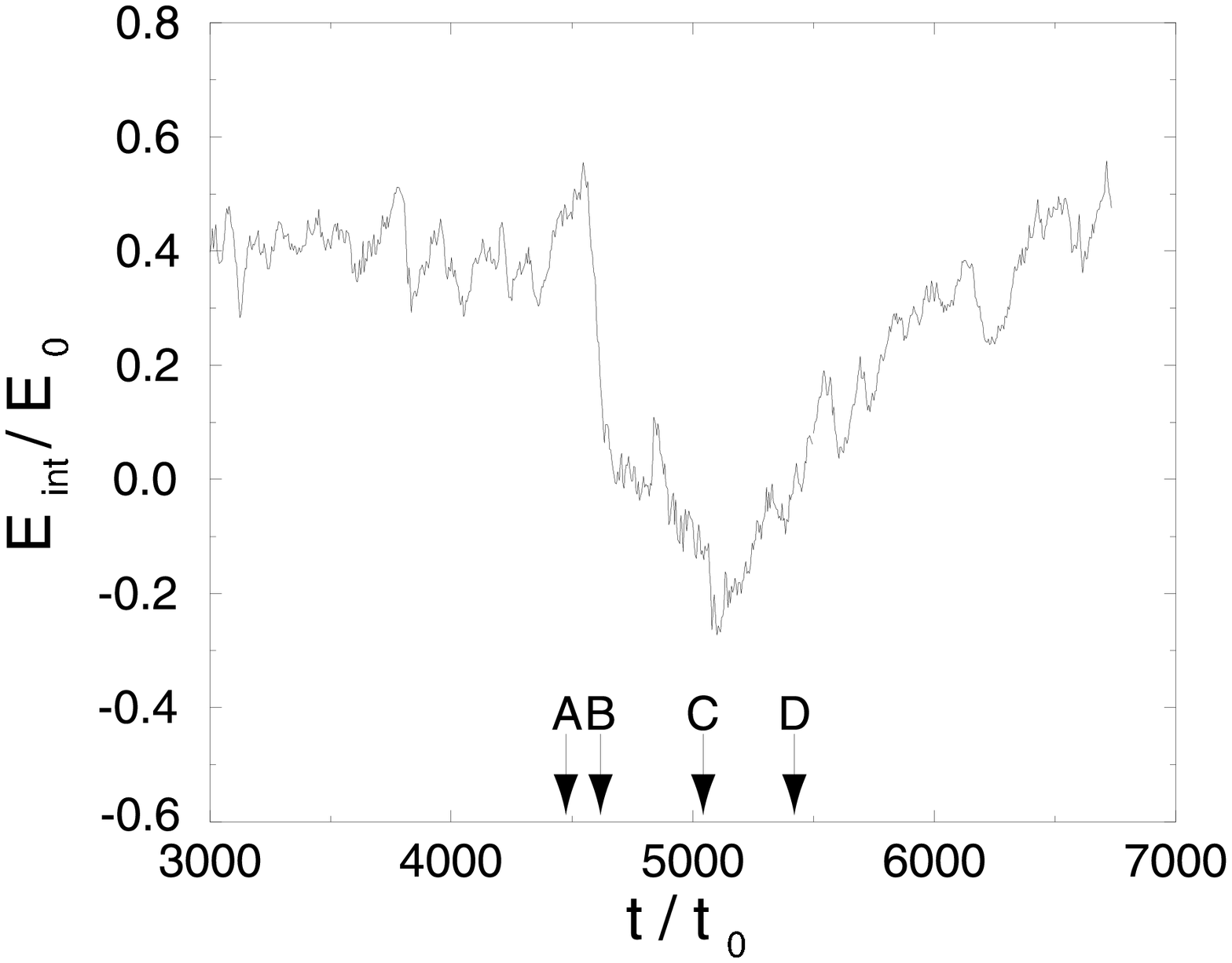}
\caption{ }
\end{figure}

\newpage

\begin{figure}[htbp]
\includegraphics[width=5.5in]{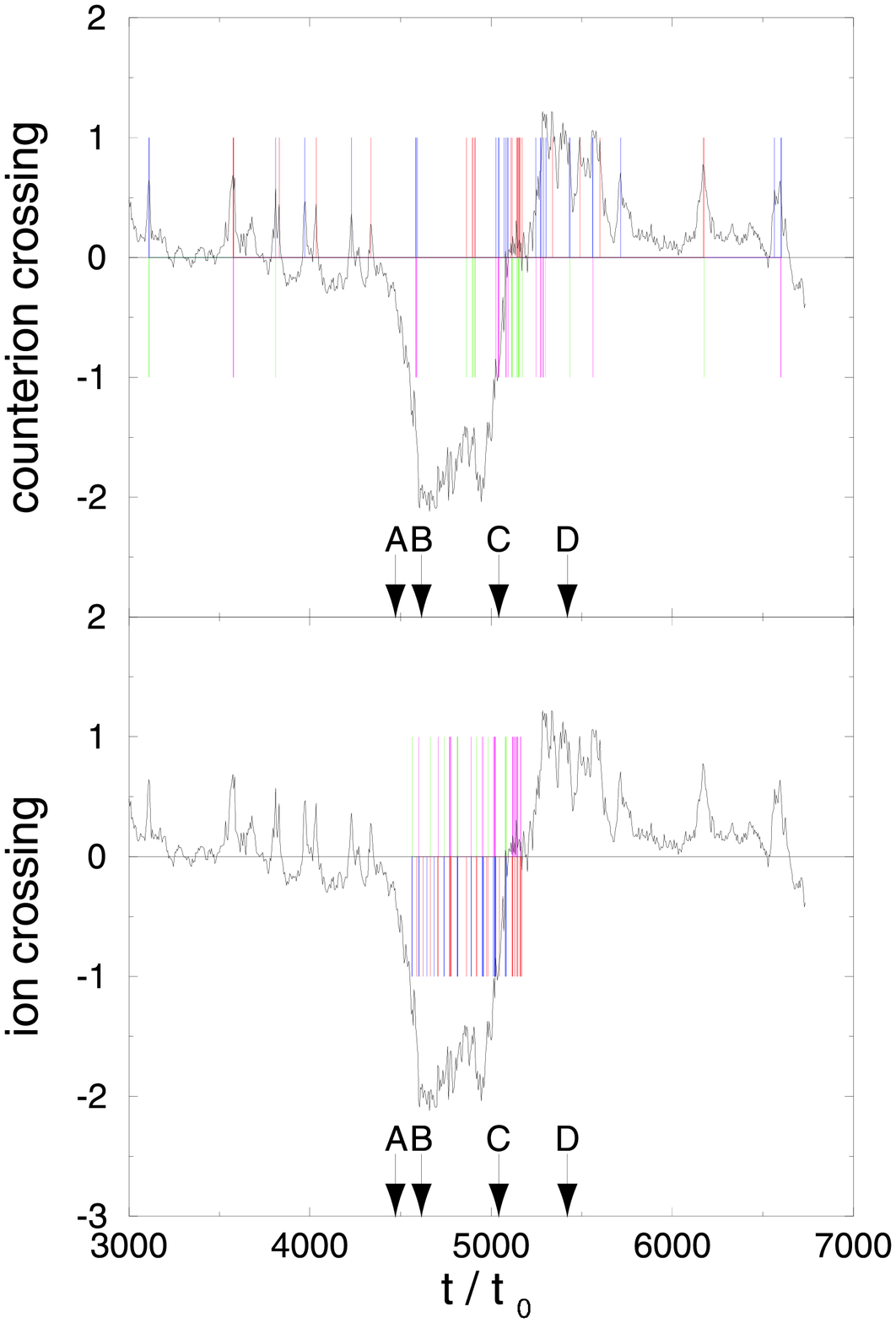}
\caption{ }
\end{figure}

\newpage

\begin{figure}[htbp]
\includegraphics[width=5.5in]{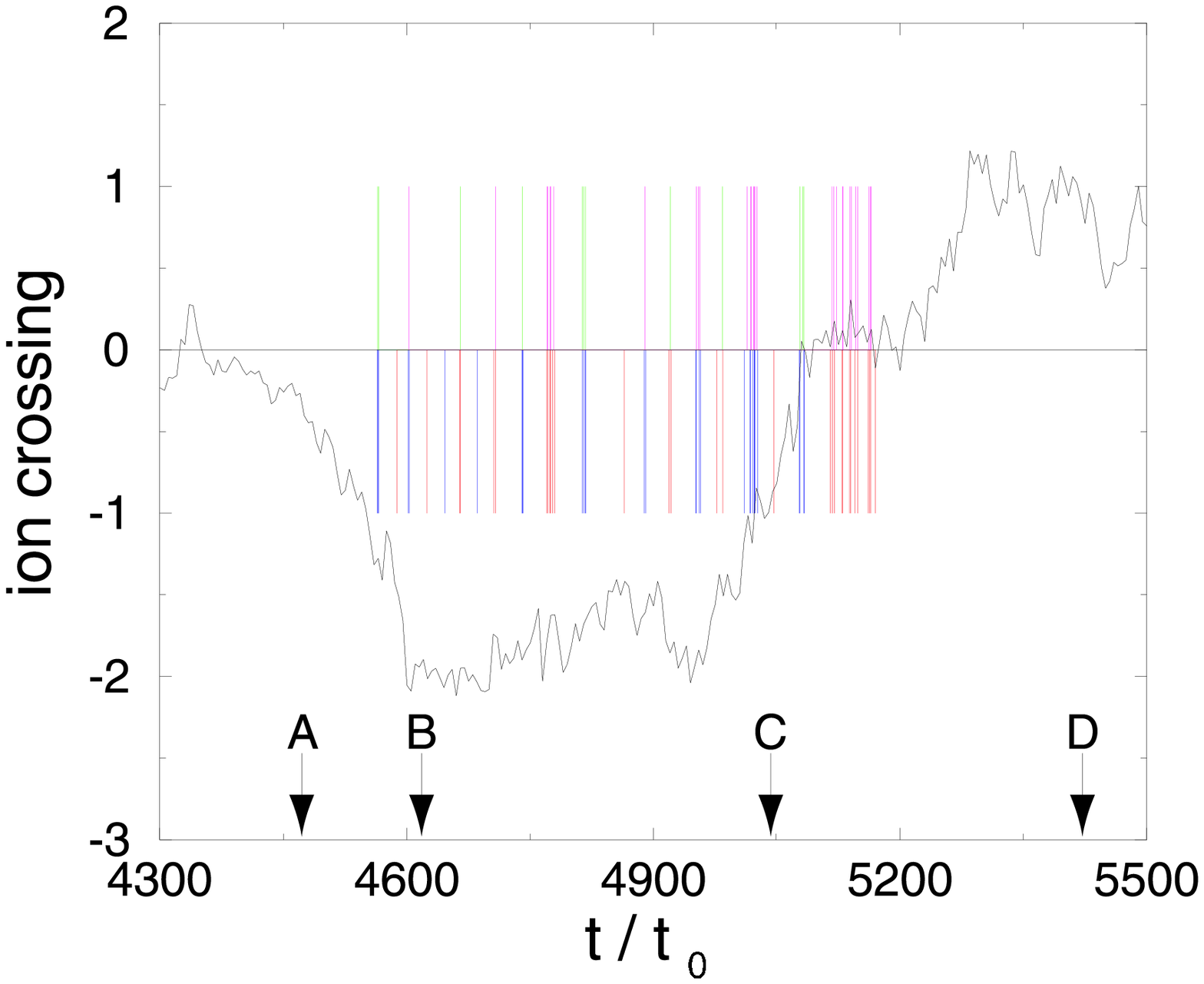}
\caption{ }
\end{figure}


\begin{thebibliography}{myarticle}

\bibitem[Ambj\"ornson et al.(2002)]{apell}
Ambj\"ornsson, T., S. \ P. \ Apell, Z. \ Konkoli, E. \ A. \ Di \ Marzio,
and J. \ J. \ Kasianowicz. 2002. Charged polymer membrane translocation.
{\it J.\ Chem. \ Phys.}\ 117:4063--4073. 

\bibitem[Berendsen et al.(1984)]{weak_coupling}
Berendsen, H. J. C., J.\ P.\ M.\ Postma, W.\ F.\ van Gunsteren,
A.\ Dinola, and J.\ R.\ Haak. 1984. Molecular dynamics with coupling
to an external bath.
{\it J.\ Chem.\ Phys.}\ 81:3684--3690.

\bibitem[Bloomfield(1996)]{bloomfield}
Bloomfield, V. A. 1996. DNA condensation.
{\it Curr.\ Opin.\ Struct.\ Biol.}\ 6:334--341.

\bibitem[Boehm(1999)]{boehm}
Boehm, R. E. 1999. Steady--state permeation rate of homopolymer chain molecules
through a pore in a barrier.
{\it Macromolecules}\ 32:7645--7654. 

\bibitem[Bokhari et al.(2002)]{shahid}
Bokhari S. H., M. A. Glaser, H. F. Jordan, Y. Lansac, 
J. R. Sauer, and B. Van Zeghbroeck. 2002.
{\it IEEE Computer Society Bioinformatics Conference}\
(Stanford University, Palo Alto, CA, August 14-16)\ 291-302. 

\bibitem[Brilliantov et al.(1998)]{brilliantov}
Brilliantov, N. V., D. \ V. \ Kuznetsov, and R. \ Klein. 1998.
Chain collapse and counterion condensation in dilute polyelectrolyte
solutions.
{\it Phys.\ Rev. \ Lett.}\ 81:1433--1436. 

\bibitem[Darden et al.(1993)]{darden}
Darden, T., D.\ York, and L.\ Pedersen. 1993. Particle mesh Ewald: an
N $\log$ N method for Ewald sums in large systems. 
{\it J.\ Chem.\ Phys.}\ 98:10089--10092.

\bibitem[Essmann et al.(1995)]{essmann}
Essmann, U., L.\ Perera, M.\ L.\ Berkowitz, T.\ Darden, H.\ Lee, and
L.\ G.\ Pedersen. 1995. A smooth particle mesh Ewald method.
{\it J.\ Chem.\ Phys.}\ 103:8577--8593. 

\bibitem[Ewald(1921)]{ewald}
Ewald, P. P. 1921. Die Berechnung optischer und elektrostatisher
Gitterpotentiale.
{\it Ann.\ Physik.}\ 64:253.


\bibitem[Foo(1997)]{foo97}
Foo, Grace M., and R. \ B.\ Pandey. 1997. 
Nonuniversal scaling and conformational crossover of polymer chains in
an electrophoretic deposition.
{\it Phys.\ Rev. \ Lett.}\ 79:2903--2906.

\bibitem[Foo(1998)]{foo98}
Foo, Grace M., and R. \ B.\ Pandey. 1998.
Electrophoresis deposition of polymer chains on an adsorbing surface
in (2+1) dimensions: conformational anisotropy and nonuniversal coverage.
{\it Phys.\ Rev. \ Lett.}\ 80:3767--3770.

\bibitem[Fuoss et al.(1959)]{fuoss}
Fuoss, R. M., F. \ Accascina,
{\it Electrolytic Conductance}\ (Interscience, New York, 1959).

\bibitem[Gronbech--Jensen et al.(1997)]{gronbech}
Gronbech--Jensen, N., R. \ J. \ Mashl, R. \ F. \ Bruinsma, and 
W. \ M. \ Gelbart. 1997. Counterion-induced attraction between rigid 
polyelectrolytes.
{\it Phys.\ Rev. \ Lett.}\ 78:2477--2480. 

\bibitem[Ha et al.(1997)]{ha}
Ha, B.-Y., and A. \ J. \ Liu. 1997. Counterion-mediated attraction between
two like--charged rods.
{\it Phys.\ Rev. \ Lett.}\ 79:1289i--1292. 


\bibitem[Henrickson et al.(2000)]{henrickson}
Henrickson, S. E., M. \ Misakian, B. \ Robertson, and J. \ J. \ Kasianowicz.
2000. Driven DNA transport into an asymmteric nanometer--scale pore.
{\it Phys. \ Rev.\ Lett.}\ 85:3057--3060. 

\bibitem[Johnson et al.(1992)]{johnson}
Johnson, J. K., J.\ A.\ Zollweg, and K.\ E.\ Gubbins. 1992. The Lennard-Jones equation of 
state revisited.
{\it Mol.\ Phys.}\ 78:591--618. 

\bibitem[Kasianowicz et al.(1996)]{branton}
Kasianowicz, J. J., E. Brandin, D. Branton, and D. W. Deamer. 1996.
Characterization of individual polynucleotide molecules using a
membrane channel.
{\it Proc.\ Nat'l.\ Acad.\ Sci.}\ USA\ 93:13770--13773.

\bibitem[Kumar et al.(2000)]{kumar}
Kumar, K. K., and K. \ L. \ Sebastian. 2000. Adsorption--assisted translocation
of a chain molecule through a pore.
{\it Phys. \ Rev.\ E}\ 62:7536--7539. 

\bibitem[Lee et al.(2001)]{lee}
Lee, S., and W. \ Sung. 2001. Coil--to--stretch transition, kink formation, and
efficient barrier crossing of a flexible chain.
{\it Phys. \ Rev.\ E}\ 63, no.\ 021115.

\bibitem[Lubensky et al.(1999)]{lubensky}
Lubensky, D. K., and D. \ R. \ Nelson. 1999. Driven polymer translocation 
through a narrow pore.
{\it Biophys.\ J.}\ 77:1824--1838.

\bibitem[Lyubartsev et al.(1995)]{lyubartsev}
Lyubartsev, A. P., and L. \ Nordenski\"old. 1995. 
Monte Carlo simulation study of polyelectrolyte properties in the presence
of multivalent polyamine ions.
{\it J.\ Polymer \ Chem. \ B}\ 101:4335--4342. 

\bibitem[Manning(1978)]{manning}
Manning, G. S. 1978. Molecular theory of polyelectrolyte solutions
with applications to electrostatic properties of polynucleotides.
{\it Q.\ Rev. \ Biophys.}\ 11:179--246.

\bibitem[Manning(1981)]{manning81}
Manning, G. S. 1981. Limiting laws and counterion condensation in
polyelectrolyte solutions. 7. Electrophoretic mobility and conductance.
{\it J.\ Phys. \ Chem.}\ 85:1506--1515.

\bibitem[Manning et al.(1994)]{ray}
Manning, G. S., and Jolly Ray. 1994. Fluctuations of counterions condensed 
on charged polymers.
{\it Langmuir}\ 10:962--966. 

\bibitem[Meller et al.(2001)]{meller}
Meller, A., L. \ Nivon, D. \ Branton. 2001.
Voltage-driven DNA translocation through a nanopore.
{\it Phys.\ Rev.\ Lett.}\ 86:3435--3438.

\bibitem[Muthukumar(1999)]{muthu}
Muthukumar, M. 1999. Polymer translocation through a hole.
{\it J.\ Chem.\ Phys.}\ 111:10371--10374.

\bibitem[Muthukumar(2001)]{muthu2}
Muthukumar, M. 2001. Translocation of a confined polymer through a hole.
{\it Phys.\ Rev.\ Lett.}\ 86:3188--3191.

\bibitem[Oosawa(1971)]{oosawa}
Oosawa, F. 
{\it Polyelectrolytes}\ (Marcel Dekker, New York, 1971).

\bibitem[Parsegian(1969)]{parsegian}
Parsegian, A. 1969. Energy of ion crossing a low dielectric membrane:
solutions to four relevant electrostatic problems.
{\it Nature}\ 221:844--846.


\bibitem[Schiessel et al.(1998)]{pincus}
Schiessel, H., and P. \ Pincus. 1998. Counterion--condensation--induced
collapse of highly charged polyelectrolytes.
{\it Macromolecules}\ 31:7953--7959.

\bibitem[Sebastian et al.(2000)]{sebastian}
Sebastian, K. L., and Alok K. \ R. \ Paul. 2000. 
Kramers problem for a polymer in a double well.
{\it Phys. \ Rev.\ E}\ 62:927--939. 


\bibitem[Slonkina et al.(2002)]{slonkina}
E.\ Slonkina, and A. \ B. \ Kolomeisky, Personal communication,
{\it arXiv:cond-mat/0209116v1}\ (5 Sep. 2002).

\bibitem[Stevens et al.(1995)]{kremer}
Stevens, M. J., and K. \ Kremer. 1995. 
The nature of flexible linear polyelectrolytes in salt free solution:
a molecular dynamics study.
{\it J.\ Chem. \ Phys.}\ 103:1669--1690. 

\bibitem[Stevens(1999)]{stevens99}
Stevens, M. J. 1999. Bundle binding in polyelectrolyte solutions.
{\it Phys.\ Rev. \ Lett.}\ 82:101--104. 

\bibitem[Sung et al.(1996)]{sung}
Sung, W., and P. \ J. \ Park. 1996. Polymer translocation through a pore
in a membrane.
{\it Phys.\ Rev.\ Lett.}\ 77:783--786. 

\bibitem[Tanaka et al.(2002)]{tanaka}
Tanaka, M., and A. \ Yu. \ Grosberg. 2002. Electrophoresis of a 
charge--inverted macroion complex: molecular--dynamics study.
{\it Eur.\ Phys. \ J. \ E}\ 7:371--379. 

\bibitem[Tinland et al.(1997)]{tinland}
Tinland, B., A. \ Pluen, J. \ Sturm, G. \ Weill. 1997.
Persistence length of single--stranded DNA.
{\it Macromolecules}\ 30:5763--5765. 

\bibitem[Toukmaji et al.(1996)]{toukmaji}
Toukmaji, A. Y., and J.\ A.\ Board Jr. 1996. Ewald summation techniques in 
perspective: a survey.
{\it Comput.\ Phys.\ Commun.}\ 95:73--92. 

\bibitem[Zimm et al.(1992)]{zimm}
Zimm, B. H., and S.\ D.\ Levene. 1992.
Problems and prospects in the theory of gel--electrophoresis of DNA.
{\it Q.\ Rev.\ Biophys.}\ 25:171--204.

\end{thebibliography}
\end{document}